\documentclass[aps,prb,amsfonts,showpacs,floats,floatfix,twocolumn,nofootinbib,byrevtex]{revtex4}
\usepackage{epsfig}
\usepackage{amsmath}
\usepackage{sidecap}
\usepackage{color}

\begin{document}
\title{Structural properties and optical response of Na clusters in
  Ne, Ar, and Kr matrices} 
\author{F. Fehrer}
\affiliation{
   Institut f\"ur Theoretische 
   Physik, Universit{\"a}t Erlangen, Staudtstrasse 7, D-91058 Erlangen,
   Germany 
}
\author{P. M. Dinh}
\affiliation{
  Laboratoire de Physique Th\'eorique, IRSAMC, CNRS, Universit{\'e} de
  Toulouse, 118 route de Narbonne F-31062 Toulouse C\'edex, France 
}
\author{P.-G. Reinhard}
\affiliation{
   Institut f\"ur Theoretische 
   Physik, Universit{\"a}t Erlangen, Staudtstrasse 7, D-91058 Erlangen,
   Germany 
}
\affiliation{
  Laboratoire de Physique Th\'eorique, IRSAMC, CNRS, Universit{\'e} de
  Toulouse, 118 route de Narbonne F-31062 Toulouse C\'edex, France 
}
\author{E. Suraud}
\affiliation{
  Laboratoire de Physique Th\'eorique, IRSAMC, CNRS, Universit{\'e} de
  Toulouse, 118 route de Narbonne F-31062 Toulouse C\'edex, France 
}
%\date{1. Draft: 28. July 2004} 
%
\begin{abstract}
We discuss the structural properties and optical response of a small
Na cluster inside rare gas (RG) matrices of Ne, Ar, or Kr atoms. The mixed
systems are described with a hierarchical model, treating the cluster
at a quantum mechanical level and the matrix atoms classically in
terms of their positions and polarizations.  We pay special attention
to the differences caused by the different matrix types. These
differences can be explained by the interplay of core repulsion
and dipole attraction in the interaction between the cluster
electrons and the RG atoms.
\end{abstract}
\pacs{36.40.Gk,36.40.Jn,36.40.Qv,36.40.Sx}  
\maketitle 

\section{introduction}

The study of the properties of clusters embedded in a matrix or
deposited on a substrate has motivated many researches since several
years~\cite{Hab94a,Hab94b}.
This setup becomes increasingly important because the
embedding/depositing simplifies the experimental handling and because
it is a generic test case for composite materials. 
As in free clusters, the doorway to (laser induced) cluster dynamics
is the optical response,
especially in metal clusters where the Mie plasmon dominates the
optical properties~\cite{Kre93}.  Let us mention as examples the
systematics of optical response in large noble-metal
clusters~\cite{Wen99a,Wen99b,Nil00,Gau01} and its dependence on the
environment~\cite{Die02}. 
The study of the optical response thus constitutes the key for
understanding the response of clusters to electromagnetic probes and it
also serves as a powerful tool for analyzing the underlying cluster
structure.

The case of inert environments is especially interesting because it
implies only moderate perturbations of cluster properties. One can
thus benefit from the well defined conditions from the surrounding
system and still access predominantly the cluster properties. 
But the theoretical modeling becomes much more involved
and the development of reliable as well as inexpensive approaches 
is still a timely task, in particular what truly dynamical
applications are concerned~\cite{Che03a,Mar03b,Rei03a}. Nonetheless,
fully 
detailed calculations have been  undertaken where details count, e.g. for the
structure of small Na clusters on NaCl~\cite{Hak96b} or the deposit
dynamics of Pd clusters on a MgO substrate~\cite{Mos02a}.
But the expense for a fully fledged quantum simulation grows huge.
These subtle models are hardly extendable to truly dynamical
situations, to larger clusters or substrates, and to systematic
explorations for broad variations of conditions.  This holds true not only
for clusters on substrates, but for all composite systems.  Thus
there exists a great manifold of approximations which aim at an
affordable compromise between reliability and expense.  One route, for
example, keeps all constituents at the same level, but simplifies the
description in terms of a microscopically founded tight-binding
approach~\cite{Wan04a,Lv05,Wan07,Wal04,Kol05,Lin05a}. At the other
extreme, one can consider all degrees of freedom as classical and
perform pure mole\-cular dynamics, as e.g. the deposition dynamics of Cu
clusters on metal~\cite{Che94} or Ar~\cite{Rat99} surface, and of Al
or Au clusters on SiO$_2$~\cite{Tak01a}. 
In between, one can take advantage 
of the very different importance or activity within the composite and thus
develop a hierarchical modeling using various levels of
approximation for the different subsystems.  Such approaches are
widely used in quantum chemistry, often called
quantum-mechanical-molecular-mechanical (QM/MM) model. They have been
applied for instance to chromophores in 
bio-molecules~\cite{Gre96a,Tap07}, surface physics~\cite{Nas01a,Inn06},
materials physics~\cite{Rub93,Kur96,Ler98,Ler00}, embedded
molecules~\cite{Sul05a} and ion channels of cell
membranes~\cite{Buc06}.  A variant 
of mixed modeling was applied in the case of a Cs atom in He
environment, the latter also considered as a quantum
system~\cite{Nak02a}.  

We are dealing here with Na clusters in rare-gas environments (Ne, Ar,
Kr). The large difference between cluster metals (reactive) and rare
gas (inert) naturally suggests a hierarchical model where the
substrate atoms are handled at a lower level of description, as
classical particles but with a dynamical polarizability.  Taking up
previous developments from~\cite{Dup96,Gro98}, we have developed, in
the spirit of QM/MM approaches, a hierarchical model for Na clusters
in contact with Ar~\cite{Ger04b} and applied it to structure, optical
response~\cite{Feh05a} and to non-linear dynamics of embedded
clusters~\cite{Feh05b}. It is the aim of this paper to present a
generalization to other types of rare gases (RG), namely Kr and Ne,
and a first comparative study of the effects of different environments
on structure and optical response.  As test cases, we consider a
Na$_8$ cluster embedded in RG clusters of various sizes. Strictly
speaking, they are mixed clusters and properties which depend on the
size of the RG system are specific to mixed clusters. But we use the
mixed systems mainly as model for Na clusters embedded in a matrix and
we thus consider rather large systems. Henceforth, we will use the
notion ``matrix'' for the RG surroundings.

\section{Model}

The model has been introduced and presented in detail 
in~\cite{Feh05c}.  However for
sake of completeness, we recall in this section the ingredients
and a few relevant formulae.

The degrees of freedom of the model are the wavefunctions of valence
electrons of the metal cluster, $\{\varphi_n({\bf r}),n=1...N_{\rm el}\}$,
the coordinates of the cluster's  Na$^+$ ion cores, 
$\{{\bf R}_I,I=1...N_{\rm ion}\}$, of the Ar atoms cores Ar$^{Q+}$),
$\{{\bf R}_a,a=1...N_{\rm Ar}\}$, and of the Ar valence clouds, 
$\{{\bf R'}_a,a=1...N_{\rm Ar}\}$.
From the given total energy, the corresponding equations of motion
are derived in a standard manner by variation. This leads to the
(time-dependent) Kohn-Sham equations for the one single-particle
wavefunctions $\varphi_n({\bf r})$ of the cluster electrons, and
Hamiltonian equations of motion for the other three degrees of
freedom, thus treated by classical mole\-cular dynamics (MD). 
For the valence cluster electrons, we use
a density functional theory at the level of the time-dependent
local-density approximation (TDLDA), augmented with an average-density
self-interaction correction (ADSIC)~\cite{Leg02}. The density of these
electrons is given naturally as defined in mean-field theories and
reads $\rho_{\rm el}({\bf r}) = \sum_n\left|\varphi_n^{\mbox{}}({\bf
  r})\right|^2$.
A RG atom is
described by two constituents with opposite charge, positive RG core
and negative RG valence cloud, which allows a correct description of
polarization dynamics. In order to avoid singularities, we associate a
smooth (Gaussian) charge distribution to 
both constituents having width $\sigma_{\rm RG}$ of the order of the p
shell "size" in RG atoms, in the spirit of~\cite{Dup96}:
\begin{equation} 
\begin{split}
  \rho_{{\rm RG},a}({\bf r}) &= 
  \frac{e \, Q}{\pi^{3/2}_{\mbox{}}\sigma_{\rm RG}^3} \times \cr
  & \hspace{-1cm} \times 
  \Big[
   \exp{\left(-\frac{({\bf r}\!-\!{\bf R}^{\mbox{}}_a)^2}{\sigma_{\rm
  Ar}^2}\right)} 
   -
   \exp{\left(-\frac{({\bf r}\!-\!{\bf R}'_a)^2}
                    {\sigma_{\rm Ar}^2}\right)}
  \Big].
%  \Big[
%   e^{-({\bf r}-{\bf R}_a)^2/\sigma_{\rm RG}^2} -
%   e^{-({\bf r}-{\bf R}'_a)^2/\sigma_{\rm RG}^2}
%  \Big].
\label{eq:RGdistri}
\end{split}
\end{equation}
The corresponding Coulomb potential exerted by the RG atoms is
related to the charge distribution (\ref{eq:RGdistri}) by the Poisson
equation, and reads:
\begin{equation} 
\begin{split}
  V^{\rm(pol)}_{{\rm RG},a}({\bf r})
  =
  e^2{Q^{\mbox{}}} 
  \Big[
  &\frac{\mbox{erf}\left(|{\bf r}\!-\!{\bf R}^{\mbox{}}_a|
          /\sigma_{\rm RG}^{\mbox{}}\right)}
        {|{\bf r}\!-\!{\bf R}^{\mbox{}}_a|} \cr
  & -
   \frac{\mbox{erf}\left(|{\bf r} \!-\! {\bf R}'_a|/\sigma_{\rm
  RG}^{\mbox{}}\right)}  {|{\bf r}\!-\!{\bf R}'_a|}
  \Big],
\label{eq:RGpolpot}
\end{split}
\end{equation}
where \mbox{$\mbox{erf}(r) = \frac{2}{\sqrt{\pi}}\int_0^r \textrm
  dx\,e^{-x^2}$} stands for the error function.
As for the Na$^+$ ions, their dynamical polarizability is
neglected and we treat them simply as charged point particles.

The total energy of the system is composed as:
\begin{equation}
  E_{\rm total}
  =
  E_{\rm Na cluster}
  +
  E_{\rm RG}
  +
  E_{\rm coupl}
  +
  E_{\rm VdW}
  \quad.
\label{eq:Etot}
\end{equation}
The energy of the Na cluster $E_{\rm Na cluster}$ consists out of
TDLDA (with SIC) for the electrons, MD for ions, and a coupling of
both by soft, local pseudo-potentials, for details
see~\cite{Kue99,Cal00,Rei03a}. 
The RG system and its coupling to the cluster are described by
\begin{widetext}
\begin{eqnarray}
\label{eq:E_RG}
  E_{\rm RG}
  &=&
  \sum_a \frac{{\bf P}_a^2}{2M_{\rm RG}} 
  +
  \sum_a \frac{{{\bf P}'_{a}}^2}{2m_{\rm RG}}
  +
  \frac{1}{2} k_{\rm RG}\left({\bf R}'_{a}\!-\!{\bf R}_{a}\right)^2
  +
  \sum_{a<a'}
  \left[
    \int \textrm d{\bf r} \, \rho_{{\rm RG},a}({\bf r})
    V^{\rm(pol)}_{{\rm RG},a'}({\bf r})
    +
    V^{\rm(core)}_{\rm RG,RG}({\bf R}_a \!-\! {\bf R}_{a'})
  \right],
\\
\label{eq:E_coupl}
  E_{\rm coupl}
  &=&
  \sum_{I,a}\left[
    V^{\rm(pol)}_{{\rm RG},a}({\bf R}_{I})
    +
    V'_{\rm Na,RG}({\bf R}_I \!-\! {\bf R}_a)
  \right]
  +
  \int \textrm d{\bf r}\rho_{\rm el}({\bf r})\sum_a \left[
    V^{\rm(pol)}_{{\rm RG},a}({\bf r})
    +
    W_{\rm el,RG}(|{\bf r} \!-\! {\bf R}_a|)
  \right],
\\
\label{eq:EvdW}
  E_{\rm VdW}
  &=&  
  \frac{e^2}{2} \sum_a \alpha_a
  \Big[
    \frac{
       \left(\int{\textrm d{\bf r} \, {\bf f}_a({\bf r}) \rho_{\rm
  el}({\bf r})}\right)^2 }{N_{\rm el}}
      - \int{\textrm d{\bf r} \, {\bf f}_a({\bf r})^2 \rho_{\rm el}({\bf
  r})} 
  \Big] \;,
  \quad
  \mbox{where} \quad 
  {\bf f}_a({\bf r})
  =
  \nabla\frac{\mbox{erf}\left(|{\bf r}\!-\!{\bf R}^{\mbox{}}_a|
          /\sigma_{\rm RG}^{\mbox{}}\right)}
        {|{\bf r}\!-\!{\bf R}^{\mbox{}}_a|}
  \quad.
\end{eqnarray}
\end{widetext}

The Van der Waals interaction between cluster electrons and RG
dipoles is written in Eq.~(\ref{eq:EvdW}) as a correlation from the
dipole excitation in the RG atom coupled with a dipole excitation in
the cluster, using the regularized dipole operator
${\bf f}_a$ corresponding to the smoothened RG charge
distributions~\cite{Gro98,Feh05c}. 

The interaction of one RG atom with the other constituents (RG
atoms, Na$^+$ ions, cluster electrons) results from the balance 
between a strong repulsive core potential that falls off exponentially
and an equally strong attraction from dipole polarizability. 
The (most important) polarization potential is described by 
a valence electron cloud oscillating against the RG core
ion. Its parameters are the effective charge of valence cloud $Q$, 
the effective mass of valence cloud $m_{\rm Ar}=Qm_{\rm el}$, 
the restoring force for dipoles $k_{\rm RG}$, and 
the width of the core and valence clouds $\sigma_{\rm RG}$.
The $Q$ and $k_{\rm RG}$ are adjusted to reproduce experimental
data on dynamical
polarizability $\alpha_D(\omega)$ of the RG atom at low frequencies,
namely the static limit 
$\alpha_D(\omega\!=\!0)$ and the second derivative
of $\alpha''_D(\omega\!''=\!0)$.
The width $\sigma_{\rm RG}$ is determined consistently such that the
restoring force from the folded Coulomb force (for small
displacements) reproduces the spring constant $k_{\rm RG}$.

The short range repulsion is provided by the various core potentials.
For the RG-RG core interaction in Eq.~(\ref{eq:E_RG}), we employ a
Lennard-Jones type potential with parameters reproducing binding
properties of bulk RG~: 
\begin{equation}
  V_{\rm RG,RG}^{\rm (core)}(R)
  = 
  e^2 A_{\rm RG} \left[
  \left( R_{\rm RG}/R \right)^{12}
 -\left( R_{\rm RG}/R \right)^{6}
  \right].
\label{eq:VRGRG}
\end{equation}
The Na-RG core potential $V'_{\rm Na,RG}$ in Eq.~(\ref{eq:E_coupl}) is
chosen according to~\cite{Rez95}, within 
properly avoiding double counting of the dipole
polarization-potential, hence the following form~:
\begin{equation}
\begin{split}
  V'_{\rm Na,RG}(R)
  =
  e^2\Bigg[
  A_{\rm Na} & \frac{e^{-\beta_{\rm Na} R}}{R} \cr
  -\
  \frac{2}{1+e^{\alpha_{\rm Na}/R}}
  & \left( \frac{\alpha_{\rm RG}}{2R^4} + \frac{C_{\rm Na,6}}{R^6} 
  + \frac{C_{\rm Na,8}}{R^8}\right)
  \Bigg] \cr
  + \, e^2\frac{\alpha_{\rm RG}}{2R^3} 
  {\bf R} \cdot \nabla_{\bf R}&
  \frac{\mbox{erf}(R/\sqrt{2}\sigma_{\rm RG})}{R} \quad.
\label{eq:VpRGNa}
\end{split}
\end{equation}
The parameters are taken from literature for Na-Ne~\cite{Lap80},
Na-Ar~\cite{Rez95,Sch03} and Na-Kr~\cite{Bru91}.
The pseudo-potential $W_{\rm el,RG}$ in Eq.~(\ref{eq:E_coupl}) for the
electron-RG core repulsion has been modeled according to the proposal
of~\cite{Dup96}~:
\begin{equation}
  W_{\rm el,RG}(r)
  =
  e^2\frac{A_{\rm el}}{1+e^{\beta_{\rm el}(r - r_{\rm el})}} \quad ,
\label{eq:VRGel}
\end{equation}
with a final slight adjustment to the 
properties of a Na-RG molecule (bond length, binding energy, and
optical excitation spectrum).
Values of atomic and dimer properties used 
are reported in table~\ref{tab:rg_dat}.

\begin{table}[htbp]
\begin{center}
\begin{tabular}{|l|l|l|l|l|l|l|}\hline
RG & \multicolumn{2}{l|}{RG-Atom} & \multicolumn{2}{l|}{Na-RG} & \multicolumn{2}{l|}{RG bulk}   \\ \hline
   &  $\alpha_{\rm RG}$ $[a_0^{-3}]$ & IP [Ry]& $d_0$ [$a_0$] &  $E_0$ [mRy] & $r_s$[$a_0$] & E$_{\rm coh}$ [mRy]\\[2pt] \hline
Ne &   2.67          & 1.585   & 10.01 &  0.0746                & 5.915 & -0.0272\\
Ar &  11.08          & 1.158   &  9.47 &  0.3793                & 7.086 & -0.1088\\
Kr &  16.79          & 1.029   &  9.29 &  0.6238                & 7.540 & -0.1497\\ \hline
\end{tabular}
\caption{
Properties of the RG atoms and Na-RG dimers for Ne, Ar, and
Kr which were used for the fine-tuning of the model.
\label{tab:rg_dat}}
\end{center}
\end{table}

\section{Modus operandi}

The numerical solution proceeds with standard methods as described in
detail in \cite{Cal00}.  The TDLDA equations for the cluster electrons
are solved on a grid in coordinate space, using a time-splitting
method for the propagation and accelerated gradient iterations for the
stationary solution. We furthermore employ the cylindrically-averaged
pseudo-potential scheme (CAPS) as an approximation for the electrons
\cite{Mon94a,Mon95a}, which is justified for the chosen embedded
Na$_8$ cluster. 
We have checked that a 2D calculation with CAPS and a full 3D
treatment of the valence electron wavefunctions both give almost 
identical optical responses.
The Na$^+$ ions as well as the RG atoms are treated in
full 3D.  The dynamics of the Na electrons is coupled to the response
of the RG dipoles. However, the ionic and atomic positions can safely
be frozen for the present study as we focus on exploring the optical
response of the embedded clusters.

To find an optimal Na+RG configuration, one starts with a fcc RG
crystal, cuts from that a given number of  closed shells, and cools the
resulting configuration for a pure RG cluster. One then carves a
cavity of 13 atoms (Ar, Kr) or 19 atoms (Ne) from the center and
places the Na$_8$ cluster into it. This mixed configuration is
re-optimized by means of successively cooled molecular dynamics for
the ions and atoms coupled to the stationary solution for the
cluster electrons.

The stationary solution of the equations of motion provides the ground
state of the mixed system and constitutes the 
initial condition for further dynamical calculations. One can 
then compute several 
observables to analyze both the statics (structural properties) and 
the dynamics. 
A global measure for
ionic and electronic cluster structure are the r.m.s. radii,
\begin{equation}
r_{I,e}= {\sqrt{\langle{(x^2+y^2+z^2)}\rangle_{I,e}}}
\label{eq:rms}
\end{equation}
where $\langle\ldots\rangle_I= \sum_I \ldots$ and $\langle\ldots\rangle_e
=\int d^3 \mathbf r \, \rho_e({\bf r}) \ldots$
Note that these quantities may also be used for characterizing dynamics, 
although they take interest mostly on very long times when 
ions and atoms actually move. This latter aspect is not directly addressed here 
and we shall thus use them only as static quantities.  
We also evaluate the insertion energy   $E_{\rm ins}$ which is defined as 
\begin{equation}
\begin{split}
  E_{\rm ins}
  =&
  E_{\rm tot}({\rm Na_8RG}_{\rm N}) 
  + E({\rm RG}_{\rm p})\cr
  &- E({\rm RG}_{\rm N+p})
  - E({\rm Na}_8) 
  \quad,
\label{eq:na_insert}
\end{split}
\end{equation}
with $p=13$ in the case of Ar and Kr, and $p=19$ for Ne.
A further useful energetic observable is the ionization potential (IP), that is
the energy required to remove one electron from the metal cluster.
We compute it as the single-particle energy of the least bound
electron which is a reliable measure in ADSIC \cite{Leg02}.

At the side of truly dynamical properties, as already emphasized, one 
should remind  the especially important role played by the optical response. 
This observable is computed in an explicitely dynamical way. 
The dynamics is initiated by an instantaneous dipole boost
of the cluster electrons. The optical response is then obtained by
spectral analysis of the emerging time-dependent dipole
signal following the strategy proposed in~\cite{Cal95a,Cal97b,Yab96}. 

\section{Results and discussion}

\begin{figure*}[htbp]
\begin{center}
	\epsfig{file=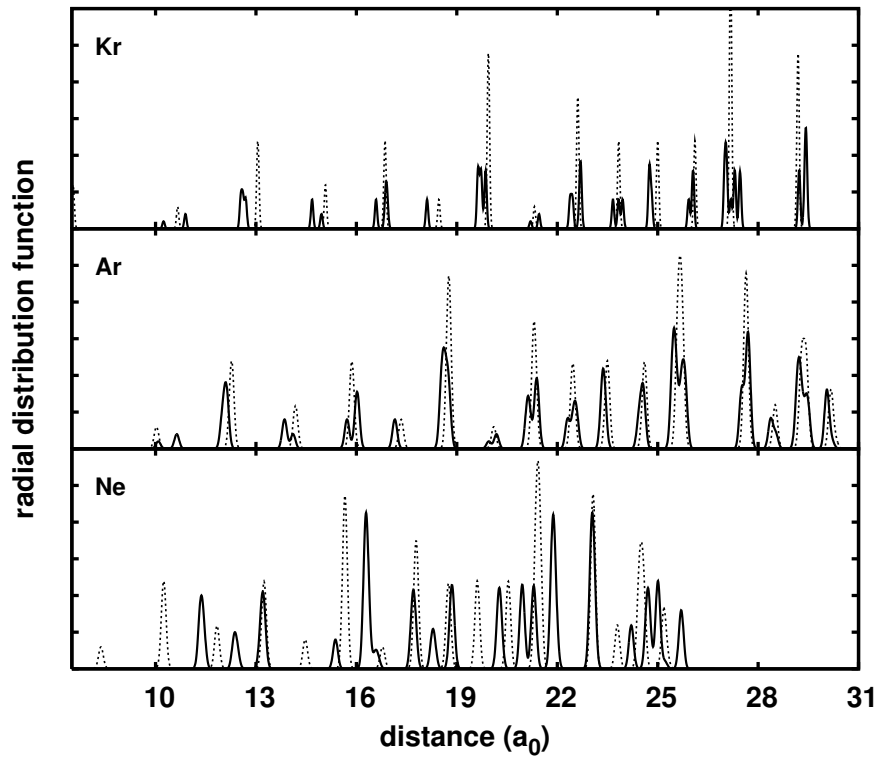,width=120mm}
	\caption{Radial distribution of RG atoms in
                 Na$_8$\@Ne$_{428}$, Na$_8$\@Ar$_{434}$, and
                 Na$_8$\@Kr$_{434}$ (full lines) compared with the
                 distributions for the pure RG cluster with
                 447 atoms before carving the cavity (dashed lines).
                  \label{fig:configs}}
\end{center}
\end{figure*}

Figure \ref{fig:configs} shows the RG structure in terms of radial
distributions of atoms. The effect of embedding is visualized by
comparing the pure RG system (dotted) with the RG distribution around
the Na$_8$ cluster (full lines). The distributions line up nicely in
radial shells.  For pure RG clusters, they remain very close to the
radial shells of the bulk fcc structure (not shown here, for the case
of Ar, see \cite{Feh05a}). The overall scale is basically given by the
bulk Wigner-Seitz radius $r_s$ given in table \ref{tab:rg_dat}. Note
that Ne has a much smaller $r_s$ resulting in denser packing as seen in
figure \ref{fig:configs}. Carving of the cavity and insertion of
Na$_8$ has only small effect in the Ar and Kr environment and mainly
for a few inner shells. For Ne, however, we see a stronger
perturbation which spreads over all atoms. The examples in figure
\ref{fig:configs} concern rather larger RG systems probably close to
bulk. The embedding effects increase with decreasing number of RG
shells, remaining small throughout for Kr and Ar, but soon destroying
any clear shell structure for Ne \cite{Fehdiss}.  The reason is that
Ne is much less bound than Ar or Kr as can be read of from the cohesion
energy in table \ref{tab:rg_dat}, and already these seemingly more
robust materials are weakly bound.

\begin{figure*}[htbp]
\begin{center}
\epsfig{file=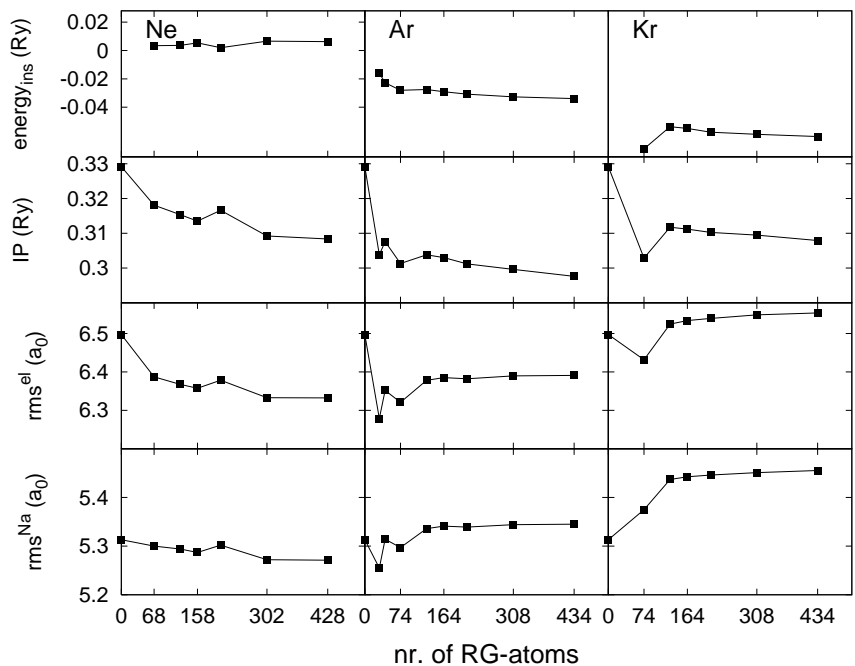,width=120mm}
\caption{Ground state observables (energies, radii) for
         Na$_8$ embedded in Ne, Ar, and Kr matrices of different size.
         \label{fig:nearkr_stat}}
\end{center}
\end{figure*}

The impact of the RG environment on metal cluster properties is
analyzed in figure \ref{fig:nearkr_stat}, which shows global
observables of the ground state configurations of Na$_8$ in the
various matrices, as a function of matrix size. 
The insertion energies in the uppermost panel indicate that Ne 
cannot finally capture Na$_8$ inside while Ar, and even more so Kr,
provide robust environments for embedded metal clusters. The 
binding increases slowly with matrix size, except for Ne where the 
polarizability is too weak to accumulate sufficient long range
attraction.
The IP (second row in figure \ref{fig:nearkr_stat}) makes a jump down
from free Na$_8$ to the embedded cases and then stabilizes with a few
fluctuations for small matrices and a faint further decrease for
larger ones. The energies are at the same scale for all RG types. The
sudden drop from free to embedded for Ar and Kr is due to the core
repulsion from the first RG shell exerted on the cluster
electrons. Adding further shells acts only indirectly by compressing
the whole matrix and thus bringing the innermost shell slightly closer
to the cluster.
Radii in Na$_8$ are shown in the two lower rows.  They vary very
little in general. There remain interesting differences in detail. The
trends with matrix size are the same for electrons and ions. However
the step from free to embedded is much different to the extent that
electrons are more compressed when embedded which is a visible effect
from core repulsion, similar as the jump in IP.
The radii decrease slightly with matrix size for Ne and increase
for Ar or Kr. This indicates that core repulsion prevails in Ne while
dipole attraction becomes more effective in Ar and Kr. 

\begin{figure}[htbp]
\begin{center}
\epsfig{file=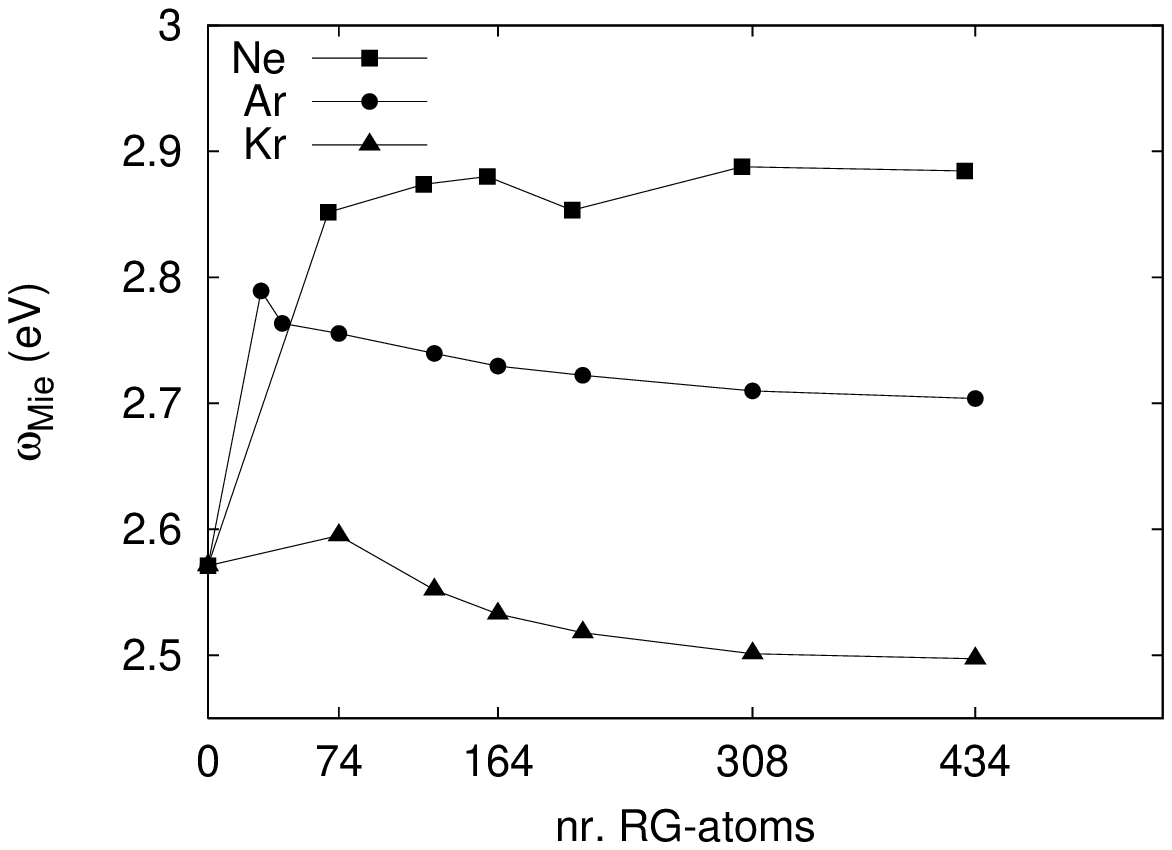,width=84mm}
\caption{Plasmon resonance energies for Na$_8$ in matrices of
  different size and RG material.
\label{fig:nearkr_plas}}
\end{center}
\end{figure}

We finally present in figure \ref{fig:nearkr_plas} the trends of the
Na$_8$ plasmon resonance peak.  The changes are generally small,
at an absolute scale at limits of our modeling (which
we estimate to about 0.1 eV uncertain). The relative trends,
however, can be taken at smaller energy scale and these 
carry several interesting aspects.
As previously discussed~\cite{Feh05a}, the position of the peak results
from a subtle cancellation 
between core repulsion and polarization effects.
The step from free to embedded clusters first produces a blue shift
because the cluster electrons feel the core repulsion from the first
layer of RG atoms. The plasmon peak moves slowly back to red with
increasing system size because each new RG shell adds to the
long-range and attractive polarization potential. 
Polarizability is also the key to the trend with RG material.
It increases with
atomic number (see column 2 of table \ref{tab:rg_dat}). The
cancellation mentioned above is thus more effective in Kr than in
Ar, and in Ar more than in
Ne. This explains the steady decrease of the peak position from Ne to
Kr and the growth of slope with atomic number.  In
fact, Ne shows no significant slope to red at all.  This happens
because added atoms compress slightly the innermost shell (see
shrinking radii with increasing system size in figure
\ref{fig:nearkr_stat}) and enhance core repulsion which, in turn,
compensates the growth of polarization effects.  Ar and Kr experience
no such compression and have anyway the stronger polarizability.

Experimental data for Na clusters embedded in rare gas
material are not yet available.  A direct comparison has thus to be
postponed. But there exist already some data on the optical
response in the somehow similar combination of Ag clusters in
RG material:
the Lausanne group has studied, among others, Ag$_8$@Ar~\cite{Fel01},
Ag$_7$@Ar,Kr,Xe~\cite{Fed93}, and Ag$_7$@Ar,Ne~\cite{Con06}
in large RG matrices
and, the Rostock group Ag$_8$ covered by small layers of Ne,
Ar, Kr and Xe (of sizes between 4 and 135 RG atoms) all
immersed in a He droplet~\cite{Die02}.

\begin{table*}
\begin{center}
\begin{tabular}{|c|c|c|c|}
\hline
Ag$_7$@ (from~\cite{Con06}) & Ag$_7$@ (from~\cite{Fed93}) 
& Ag$_8$@ (from~\cite{Fel01}) & Ag$_8$@ (from~\cite{Die02})\\
\hline
&&&  He$_{\rm drop} \rightarrow$ Ne$_{135}$: 0.009\\
\cline{1-1} \cline{4-4}
Ne $\rightarrow$ Ar: $-0.1$ &&&\\
\cline{1-1} \cline{3-4}
&& He$_{\rm drop} \rightarrow$ Ar: $-0.03$ & 
He$_{\rm drop} \rightarrow$ Ar$_{24}$: $-0.01$ \\
\cline{2-4}
& Ar $\rightarrow$ Kr: $-0.06\ /-0.09$ &&\\
\cline{2-2} \cline{4-4}
&&& He$_{\rm drop} \rightarrow$ Kr$_{25}$: $-0.04$\\
\cline{4-4}
&&& He$_{\rm drop} \rightarrow$ Xe$_{15}$: $-0.08$\\
\hline
\end{tabular}
\end{center}
\caption{
\label{tab:plasmon}
Shift of plasmon peak position, in eV, going from one environment to
another, as indicated. A plus (minus) sign stands for a blue (red)
shift.
}
\end{table*}

Table~\ref{tab:plasmon} summarizes
the results in terms of shifts
of the plasmon peak position with changing RG material.
A direct comparison with  free neutral Ag clusters is not available
because these are hard to handle experimentally.
As for experiments
performed in~\cite{Die02}, Ag$_8$ is embedded in 
very small RG matrices, and the whole system itself embedded in a
He droplet, for reasons of better handling.
It is usually claimed that the helium environment interacts very
faintly with the embedded system and can then be considered
practically as a vacuum. However experimental data~\cite{Kin95,Kin96}
and DFT calculations~\cite{Nak02a} show that the presence of He around
Cs atoms produces a blue shift of the plasmon peak.
It is thus likely that some small blue shift exists also for
Ag clusters directly in He droplets. Therefore a comparison with truly free
clusters is excluded. On the contrary, the influence of He around the
RG layers in the experiments of~\cite{Die02} 
is most probably negligible.
The effect of the He droplet is indeed strongly shielded by the RG
layers both spatially and energetically. Remind that a typical RG-He
bond has a length of about 6--7 $a_0$~\cite{Hub79} and an energy of a
few meV~\cite{Tan86}, comparable to the metal-RG
bonds~\cite{Anc95,Jak97,Mel02} (see fourth column of
table~\ref{tab:rg_dat}). But once coated with the RG layers, the metal
cluster lies typically twice farther away from the He droplet than in
the case without RG layers, whence a vanishingly small residual
interaction between the metal cluster and the He droplet.
Thus the relative shifts that we observe fby changing RG material are
most probably reliable. Finally, a word of caution is in order. 
The above mentioned experimental measurements and our calculations on
embedded Na clusters~\cite{Feh06a} report a broadened, often even
fragmented, peak  with a width somewhat larger than the shifts we are
looking at. The comparison is thus at the edge of experimental and
theoretical resolution.

The data in table~\ref{tab:plasmon} agree with our
theoretical results in that all shifts are very small. Inert
environment turns out to be indeed inert with respect to the plasmon
peak position.
Looking in more detail at the relative shifts, one can read for the step
from Ne to Ar a red shift of 0.1 eV for Ag$_7$ in large
matrices~\cite{Con06} and of 0.02 eV for Ag$_8$ in small
matrices~\cite{Die02}.  The step from Ar to Kr yields 0.06--0.09 eV
for Ag$_7$ in large systems~\cite{Fed93} and 0.03 eV for Ag$_8$ in
small systems~\cite{Die02}.  The experiments with small RG layers in a
He droplet thus yield generally smaller red shifts. This also holds 
for the step from pure He to Ar environment. This is probably
explained by the size of the matrix, as we see from our results in
figure \ref{fig:nearkr_plas} a slow but steady move towards
red (however, systematic errors by comparing two very different
experiments cannot safely be excluded). Our results for embedded Na
clusters shown in figure \ref{fig:nearkr_plas} show a red shift of
0.1--0.2 eV for the step from Ne to Ar and of 0.15--0.2 eV for Ar to
Kr, both growing with increasing system size.
They confirm all trends seen in experiment but are generally half an order
of magnitude larger.
This is probably due to the smaller Wigner-Seitz radius of Ag
(3 a$_0$ instead of 4 a$_0$ for Na) which means
that Ag structures are much more compact than Na ones and thus couple
less strongly to the matrix because both fill the same RG
cavity. And this, in turn, produces smaller shifts.
In order to check that argument, we have simulated an embedded
``Ag$_8$'' cluster simply by rescaling the ionic positions of the
Na$_8$ by the ratio of Wigner-Seitz radii, that is 3/4, and by
reoptimizing the RG positions before the calculations of the optical
response of this pseudo Ag$_8$. We find the same trends as
with Na$_8$ in figure \ref{fig:nearkr_plas}, but indeed
reduced by a factor of three. 

Besides, previous theoretical calculations were performed within TDLDA
and a jellium model, coupling to the RG materials in terms of
a static dielectric medium with dielectric constant $\varepsilon$ for
different metal clusters, namely K~\cite{Rub93}, Na and
Al~\cite{Kur96}, and Ag~\cite{Ler98}. They always yield 
red shifts which, moreover, increase with increasing $\varepsilon$.
Indeed, the ingredients added in these models can only produce
long-range polarization effects and thus can only give a red shift.
Our model contains as new component,  the RG short-range and
repulsive core potentials which generate a blue shift of the plasmon
peak.
A clear experimental assessment  would yet require 
a comparison with truly free neutral metal clusters.

\section{Conclusions}

We have discussed in this paper static properties and optical response
of a small Na cluster embedded in various rare gas (RG) matrices. 
For this purpose, we used a recently introduced hierarchical approach
combining a fully detailed quantum-mechanical description of the
cluster with a classical modeling for the RG environment and its
interactions with the cluster. We have studied effects from embedding
in RG environment with up to 434 atoms with particular emphasis
on the change with RG material. Various observables were considered.
The insertion energy yields stable embedding for Ar and Kr but not for
Ne within the considered system sizes. 
The IP behaves the same in all three materials: It drops from free to
embedded and stays nearly constant for all matrix sizes. 
The electronic and ionic radii of the Na cluster change very little;
the most noteworthy effect is here a slight compression of the
electron cloud through embedding.
What optical response is concerned, we studied the effect of
the RG matrices on the position of the surface plasmon peak in the
metal cluster. The net shift from free to embedded clusters
is very small due to a near cancellation of the
blue shift from core repulsion with the red shift
from dipole polarization in the interaction with the RG
atoms. The polarization increases with the RG atomic weight and thus
final peak position goes from blue shift to red shift on the way from
Ne over Ar to Kr.  The polarization effect also increases
with the size of the RG matrix which produces a slow and steady trend
to red with increasing system size.
Comparison with experimental data on embedded Ag clusters confirms
these trends and orders of magnitude.  
Having explored the basic observables of structure and optical
response, the way is open to applications in truly dynamical
scenarios. The large difference in atomic masses of the RG
will then play a crucial role and yield interesting effects. Work in
that direction is under progress.

\acknowledgments
This work was supported by the DFG (RE 322/10-1),
%nr.  04670PG,
 the CNRS Programme "Mat\'eriaux"
  (CPR-ISMIR), Institut Universitaire de France, and the
  Humboldt foundation.% and a Gay-Lussac prize.

\bibliographystyle{apsrev}
\bibliography{cluster,KrArNe_opt,add-KrArNe}

\end{document}